\documentclass[aps,prd,twocolumn,amssymb,amsmath,showpacs,a4paper,superscriptaddress]{revtex4-1}
\usepackage{graphicx}
\usepackage{amsfonts}
\usepackage{amsmath}
\usepackage{hyperref}
\usepackage{dcolumn}% Align table columns on decimal point
\usepackage{bm}

\usepackage{latexsym}
\usepackage{amsfonts}
\usepackage{psfrag}
\usepackage{graphicx}
\usepackage{amssymb,amsmath} %,epsfig
\usepackage{appendix}

\usepackage{color}
\definecolor{myOrange}{rgb}{1,0.5,0.1}
\definecolor{myRed}{rgb}{0.8,0.1,0.1}
\definecolor{myGreen}{rgb}{0.7,0,0.8}
\definecolor{myGray}{rgb}{0.6,0.6,0.6}

\definecolor{light-gray}{gray}{0.95}

\begin{document}

\setlength{\pdfpagewidth}{8.5in}
\setlength{\pdfpageheight}{11in}

\title{Superradiant scattering of orbital angular momentum beams}

\author{Cisco Gooding}%\email{cgooding@physics.ubc.ca}
\affiliation{
	School of Mathematical Sciences, 
	University of Nottingham, UK
	}
\affiliation{
  Department of Physics \& Astronomy,
  University of British Columbia, Canada
}
\author{Silke Weinfurtner}%\email{Silke.Weinfurtner@nottingham.ac.uk}
\affiliation{
	School of Mathematical Sciences, 
	University of Nottingham, UK
}
\affiliation{
	Centre for the Mathematics and Theoretical Physics of Quantum Non-Equilibrium Systems, 
	University of Nottingham, UK
}
\author{William G. Unruh}%\email{unruh@physics.ubc.ca}
\affiliation{
  Department of Physics \& Astronomy,
  University of British Columbia, Canada
}
\affiliation{Hagler IAS, IQSE, Texas A\&M, College Station, TX, 77843-4242, USA}
\date{\today}

\begin{abstract}
We consider the wave-structure coupling between an acoustic orbital angular momentum beam and a rapidly rotating disk, and present a new configuration exhibiting the wave amplification effect known as rotational superradiance. While initially envisioned in terms of the scattering of an incident wave directed perpendicular to an object's rotation axis, we demonstrate in the context of acousto-mechanics that superradiant amplification can also occur with a vortex beam directed parallel to the rotation axis. 
We propose two different experimental routes: one must either work with rotations high enough that the tangential velocity at the outer edge of the disk exceeds the speed of sound, or use evanescent sound waves. We argue that the latter possibility is more promising, and provides the opportunity to probe a previously unexamined parameter regime in the acoustics of rotating porous media.
\end{abstract}

\maketitle

\section{Introduction}
An orbital angular momentum (OAM) beam is a traveling wave with angular momentum in the direction of its propagation. Rather than angular momentum associated with spin degrees of freedom, OAM beams result from spatial wave distributions that have a helical structure. OAM beams can be formed in nearly any effective field theory with propagating modes, such as electromagnetism (i.e. light beams \cite{Allen92}) and fluid dynamics (i.e. sound beams in air or water), and as such have facilitated the widespread application of rotation for science and industry alike. Despite their interest and applicability there is still much to be learned about how OAM beams interact with rapidly rotating objects. In this work, we present a novel approach demonstrating the amplification of OAM beams from a rapidly rotating disk. 

Our proposal complements the well-known configuration used to describe rotational superradiance, a celebrated theoretical effect that can occur in gravitational \cite{Penrose71,Bekenstein73}, electromagnetic \cite{Zeldovich71,Zeldovich72}, and fluid dynamical systems \cite{Silke2016}. Theoretical work on rotational superradiance began in the $1970$s, though the effect has eluded experimental confirmation until only recently. The first detection of rotational superradiance was achieved for surface waves interacting with a vortex flow~\cite{Silke2017}. One of the major difficulties inherent to superradiance experiments stems from geometry: given a rotating, partially absorbing object, the standard approach follows Zel'Dovich's original conception, which involves scattering a wave directed perpendicular to the object's rotation axis. Consequently, the incident wave scatters in a continuum of directions, making detection and subsequent analysis of the scattered wave problematic.

Instead of perpendicular, we propose to direct an incident wave (carrying angular momentum) parallel to the object's rotation axis. Hence, within this configuration it is possible to utilise the arguably advantageous OAM beams to probe for superradiance. In cylindrical coordinates $(r,\theta,z)$ with the $z$-axis aligned with the object's angular momentum, our new configuration amounts to scattering in the $z$-direction, as opposed to in the $(r,\theta)$ plane. We therefore reduce the \emph{effective} dimensionality of the scattering setup, resulting in a conceptual as well as \emph{practical} simplification that admits a more transparent theoretical description. The simplifications our approach brings with it broaden the range of possible applications. 

Below we will apply our new framework for superradiance to the fluid dynamical model of acoustic vortex beams. Taking advantage of the simplicity of theoretical description that follows from aligning the directions of the incident and scattered waves, we obtain a new expression for the surface impedance at the fluid-disk interface, and  make direct comparisons with previous fluid dynamical approaches. We determine how the derived impedance condition characterizes superradiance in the system, and demonstrate that superradiance can indeed occur in our proposed setup. We also prove that in the superradiant regime, one cannot both operate the disk with solely subsonic tangential velocities and use propagating modes for the wave scattering. One must therefore either rotate the disk fast enough for part of it to travel supersonically, or else work with evanescent waves.
We then focus on the experimental implementation of our proposal for sound waves in air. For simplicity of presentation, we restrict our attention to a more idealized impedance condition, and calculate the amplification spectra for experimentally feasible parameters. This is followed by a detailed discussion of vortex beam generation and detection of the scattered wave. By taking all of the above into account we argue that the implementation of our proposal is within experimental reach.
Finally, we comment on the applicability of our proposal to electromagnetic systems, which could potentially allow quantum aspects of superradiance to be observed.

\section{Theory}

\begin{figure}[t!]
  \includegraphics[width=0.95\columnwidth]{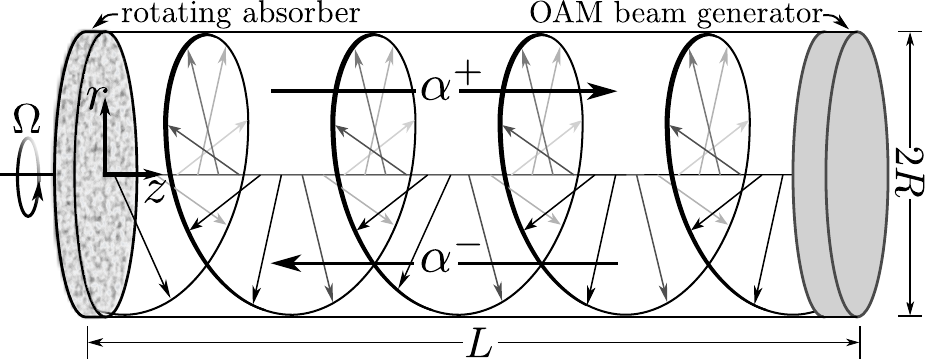}
  \caption{Schematic representation of the scattering setup.}
	\label{fig:Schem}
\end{figure}

Our system of interest is a barotropic fluid bounded by a cylindrical tube of inner radius $R$. At one end of the tube a coaxial rotating disk composed of sound-absorbing material is inserted. The disk is aligned with a cylindrical coordinate system $(r,\theta,z)$ such that the surface of the disk is located at $z=0$. At the opposite end of the tube (which is pointing at the positive $z$-direction) acoustic waves are produced, and are guided through the cylinder towards the fluid-disk interface, as depicted in Figure~\ref{fig:Schem}. Away from the disk we assume a uniform background density $\rho_0$ and vanishing background flow $\bm{u}_0$. The acoustic velocity $\bm{u}$ is then irrotational, and can be expressed in terms of a potential $\Phi$ as $\bm{u}=\nabla \Phi$. 

For a monochromatic sound wave with frequency $\omega$ and total wavenumber $k$, we can write the acoustic pressure as $p=c^2 \rho=i\omega\rho_0\Phi$, with $c=\omega/k=\sqrt{B_a/\rho_0}$ the phase speed and $B_a$ the (adiabatic) bulk modulus of the fluid. The velocity potential for such a monochromatic sound wave obeys the Helmholtz equation $\nabla^2\Phi+k^2 \Phi=0$. 

The fluid is enclosed by the tube at $r=R$, so we enforce the impermeability boundary condition,
\begin{equation}\label{BCR}
\bm{u}\cdot\hat{\bm{r}}|_{r=R}=\partial_r \Phi|_{r=R}=0.
\end{equation}
At the surface of the disk, determination of the boundary condition is less straightforward. To absorb sound, we assume a disk made of a porous/fibrous material. There is a large amount of literature devoted to the dynamics of fluids in saturated porous materials (see \cite{ConvectionPorousMedia} and references therein), though motion of the porous frame is neglected in most work. A notable exception is Auriault's analysis of acoustic waves in the saturating fluid coupled with elastic waves in deformable, poroelastic media undergoing steady rotation at an angular velocity $\bm{\Omega}$~\cite{AcoustRotDeformPorous}. Auriault uses a technique known as homogenization, which involves averaging over a representative volume element (RVE) containing sufficient pores to allow for a coarse-grained description at the macroscopic scale \cite{Homogenization}. We follow Auriault's approach to characterize the fluid-structure interaction at the disk surface, but since we ultimately wish to implement our proposal with sound waves in air, the frame can be treated as rigid: the bulk moduli of most porous materials are much larger than the bulk modulus of air, so elastic waves in the frame are not excited by acoustic waves in the fluid \cite{Porous}.

One then expects a macroscopic description of the acoustics in terms of an effective fluid propagating on a rotating background. Defining features of the effective fluid are determined by the extent of drag and other interactions with the rotating porous frame.
Within nonrotating anisotropic rigid-framed porous media, quasi-stationary fluid motion obeys Darcy's law, 
\begin{equation}\label{DarcyLett}
\bm{v}=-\frac{1}{\mu}\bm{K}(\omega)\cdot\nabla p,
\end{equation}
where $\mu$ is the fluid viscosity, $\bm{K}(\omega)$ is the permeability tensor, and the filtration velocity $\bm{v}$ (also known as the Darcy velocity) represents the pore-scale fluid velocity averaged over a RVE~\cite{DarcyOG,DynamPorousFluids}.
Isotropy at the macroscopic level corresponds to a scalar permeability, multiplied by the unit tensor. It will become apparent that linearity between the pressure gradient and the filtration velocity is still maintained when the disk undergoes a rigid rotation with constant angular velocity $\bm{\Omega}$. The tensor defining the linearity between $\nabla p$ and $\bm{v}$ can still be interpreted as a (dynamic) permeability $\bm{K}$, but unlike the nonrotating case, $\bm{K}$ is not symmetric, and depends on $\bm{\Omega}$. 

To understand how our system behaves at the fluid-disk interface, we take the rigid-frame limit of Auriault's equations ($101$)-($104$) \cite{AcoustRotDeformPorous}, which corresponds to setting the displacement of the solid frame to zero (with respect to the steady rotation at angular velocity $\bm{\Omega}$). One then finds the independent equations
\begin{equation}
\nabla\cdot\left(-i\bar{\omega}\varphi\bm{U}_f\right)=\frac{i\bar{\omega}\varphi\beta p}{\rho_0 c^2} \label{AurComp}
\end{equation}
and
\begin{equation}
-i\bar{\omega}\varphi \bm{U}_f=-\frac{1}{\mu}\bm{K}(\bar{\omega},\bm{\Omega})\nabla p, \label{AurDarcy}
\end{equation}
where $\bm{U}_f$ is the acoustic fluid displacement, $\varphi$ is the porosity, $\mu$ is the dynamic viscosity of the fluid, and $\beta$ is the dimensionless fluid compressibility. The permeability tensor $\bm{K}$ depends on the co-rotating acoustic frequency $\bar{\omega}=\omega-m\Omega$, as well as the angular velocity vector $\bm{\Omega}$, though the explicit functional form given by Auriault applies only to an idealized and very particular situation. We now provide a more general expression for the permeability, resulting in a model capable of describing the scattering of acoustic vortex beams described in our experimental proposal.

At the pore scale, the local dynamics for acoustic disturbances in a viscous, saturating fluid with respect to a corotating coordinate system are given by
\begin{equation}
\mu \nabla^2 \bm{u}_f-\nabla p=\rho \left(\partial_t\bm{u}_f+\bm{\Omega}\times\left(\bm{\Omega}\times\bm{U}_f\right)+2\bm{\Omega}\times\bm{u}_f\right), \label{PoreV}
\end{equation}
where $\bm{u}_f=\partial_t\bm{U}_f$ is the acoustic fluid velocity associated with the displacement $\bm{U}_f$. Since the pore scale is much smaller than the acoustic scale, the fluid density $\rho$ is taken to be constant here, and compressibility of the fluid is neglected. After obtaining an explicit permeability tensor, we will return to the description defined by (\ref{AurComp}) and (\ref{AurDarcy}), which takes into account compressibility.

Taking an intrinsic average of (\ref{PoreV}) over a representative volume element (RVE) containing many pores, one concludes that a linear relationship exists between $\nabla p$ and the intrinsic average fluid velocity $\langle \bm{u}_f \rangle$:
\begin{equation}
\nabla p = -\bm{H}(\bar{\omega},\bm{\Omega})\langle \bm{u}_f \rangle+\mathcal{O}\left(l/L\right),
\end{equation}  
where $l$ is the characteristic pore scale, $L$ is the macroscopic scale, and $\bm{H}=\mu \bm{K}^{-1}$ \cite{AcoustRotDeformPorous}. Auriault notes a characteristic frequency $\omega_c=\mu/\rho_0 l^2$, and distinguishes between the low frequency regime $\bar{\omega}<\omega_c$, where viscous effects are dominant, and the high frequency regime $\bar{\omega}>\omega_c$, where inertial effects are dominant.

In the high frequency regime, one can neglect the viscous term in (\ref{PoreV}). Straightforward averaging of the resulting equation then leads to the lowest-order expression
\begin{equation}\label{Tens}
-\varphi\nabla p=\rho_0\left[-i\bar{\omega} \langle \bm{u}_f \rangle-\bm{\Omega}\times\left(\bm{\Omega}\times\frac{\langle \bm{u}_f \rangle}{i\bar{\omega}}\right)+2\bm{\Omega}\times\langle \bm{u}_f \rangle\right],
\end{equation}
implying that $\bm{H}=\frac{-i\bar{\omega}\rho_0}{\varphi}\bm{A}$, where $\bm{A}$ has components
\begin{equation}
A_{ij}=\left(1+\frac{\Omega^2}{\bar{\omega}^2}\right)\delta_{ij}-\frac{1}{\bar{\omega}^2}\Omega_i\Omega_j-\frac{2i}{\bar{\omega}}\epsilon_{ijn}\Omega_{n}.\label{Inertial}
\end{equation} 
The real part of $\bm{H}$ relates to dissipation, whereas the imaginary part has inertial origins. One can see from (\ref{Inertial}) that the high frequency regime has a primarily imaginary $\bm{H}$, with real contributions coming solely from the Coriolis term.

At this point, Auriault proceeds to determine $\bm{K}(\bar{\omega},\bm{\Omega})$ for a particular idealized pore geometry composed of plane fissures, by solving the boundary value problem at the pore scale, and averaging the solution. Pore geometries one encounters in practice are often too complicated for the boundary value problem at the pore scale to be tractable, so we will instead observe from the linearity of (\ref{PoreV}) that $\bm{H}$ can be written in the form
\begin{equation}\label{HTotal}
\bm{H}(\bar{\omega},\bm{\Omega})=\bm{H}^0-\frac{i\bar{\omega}\rho_0}{\varphi}\bm{A}(\bar{\omega},\bm{\Omega}),
\end{equation} 
where $\bm{H}^0$ arises due to viscosity. For a macroscopically isotropic medium, neglecting the inertial terms leads to $\bm{H}=\bm{H}^0=\frac{\mu}{K_0}\bm{\delta}$, with $\bm{\delta}$ denoting the unit tensor. This case represents the quasi-stationary, isotropic form of Darcy's filtration law, usually expressed as
\begin{equation}\label{DarcyScalar}
\bm{v}=-\frac{K_0}{\mu}\nabla p,
\end{equation}
with $\bm{v}=\langle \bm{u}_f\rangle$ being the filtration velocity. The quantity $K_0$ can be interpreted as the \emph{static} permeability of the disk. A related proposal to detect superradiant gain in an acoustic system implicitly assumes that (\ref{DarcyScalar}) holds in rotating media as well \cite{Faccio}. For rapidly rotating media, it is more appropriate to incorporate the tensorial contributions to the permeability implied by (\ref{Tens}).

Including both viscous and inertial terms in $\bm{H}\equiv\mu\bm{K}^{-1}$, we can obtain a macroscopic description of acoustics in the rotating porous disk:
\begin{equation}\label{ExplicitH}
\bm{H}(\bar{\omega},\bm{\Omega})=\frac{\mu}{K_0}\bm{\delta}-\frac{i\bar{\omega}\rho_0}{\varphi}\bm{A}(\bar{\omega},\bm{\Omega}),
\end{equation}
with $\bm{A}$ defined by (\ref{Inertial}). The components of the resulting generalized Darcy law $\nabla p=-\bm{H}(\bar{\omega},\bm{\Omega})\bm{v}$ can be expressed in co-rotating cylindrical coordinates as
\begin{align}\label{AcoustPr}
\partial_r p = \frac{i\bar{\omega}\rho_0}{\varphi}\left[\left(1+\frac{\Omega^2}{\bar{\omega}^2}\right)v_r-2i\frac{\Omega}{\bar{\omega}}v_\theta\right]-\frac{\mu}{K_0}v_r, \\
\frac{1}{r}\partial_\theta p= \frac{i\bar{\omega} \rho_0}{\varphi}\left[\left(1+\frac{\Omega^2}{\bar{\omega}^2}\right)v_\theta + 2i\frac{\Omega}{\bar{\omega}}v_r\right]-\frac{\mu}{K_0}v_\theta,\\
\partial_z p = \left(\frac{i\bar{\omega} \rho_0}{\varphi}-\frac{\mu}{K_0}\right)v_z,\label{AxialPr}
\end{align}
and the volume-balance equation (\ref{AurComp}) can be written as
\begin{equation}\label{VolBalance}
- \beta\left(\frac{\bar{\omega}}{c}\right)^2 p =\frac{i\bar{\omega}\rho_0}{\varphi}\left[\frac{1}{r}\partial_r\left(r v_r\right)+\frac{1}{r}\partial_\theta v_\theta +\partial_z v_z\right]. 
\end{equation}
For economy of notation, we omit overbars on the co-rotating angular coordinate $\bar{\theta}$ when it appears in a subscript.

The coupled equations (\ref{AcoustPr})-(\ref{VolBalance}) can be reduced to a single acoustic equation expressed solely in terms of the pressure by inserting (\ref{AurDarcy}) into (\ref{AurComp}):
\begin{equation}\label{PressDarcy}
\beta\left(\frac{\bar{\omega}}{c}\right)^2 p = \frac{i\bar{\omega} \rho_0}{\mu\varphi}\nabla\cdot\left[\bm{K}\nabla p\right].
\end{equation}
The antisymmetric parts of $\bm{K}$ cancel out in (\ref{PressDarcy}) due to symmetry, leaving the diagonal components as the only contributors. Explicit expressions for the cylindrical components of $\bm{K}$ can be found by inverting the matrix representation of (\ref{ExplicitH}) in the $(r,\bar{\theta},z)$ basis. 

Instead of working with the permeability directly, we will follow the description of the rigid-frame limit of Biot's theory, in which sound propagation in the porous medium is fully characterized by the compressibility $\beta(\bar{\omega})$ and the (tensor-valued) tortuosity $\bm{\alpha}(\bar{\omega})=(i\varphi \mu/\bar{\omega} \rho_0)\bm{K}^{-1}$. The compressibility $\beta$ does not vary significantly in the parameter range relevant to our experimental proposal, so we will set it to unity. The pressure equation (\ref{PressDarcy}) then becomes
\begin{equation}\label{TortPress}
- \left(\frac{\bar{\omega}}{c}\right)^2 p = \nabla\cdot\left[\bm{\alpha}^{-1}\nabla p\right].
\end{equation} 
The corresponding cancellation of off-diagonal elements for $\bm{\alpha}$ implies (\ref{TortPress}) reduces to
\begin{equation}\label{TortPressDiff}
-\left(\frac{\bar{\omega}}{c}\right)^2 p=\alpha^{-1}_{rr}\frac{1}{r}\partial_r\left(r\partial_r p\right) + \alpha^{-1}_{\theta\theta}\partial_\theta^2 p+\alpha^{-1}_{zz} \partial_z^2 p.
\end{equation}
Isotropy in the transverse ($\{r,\bar{\theta}\}$) plane implies $\alpha^{-1}_{rr}=\alpha^{-1}_{\theta\theta}$. Using the ansatz $p\sim J_{\bar{m}}(\bar{k}_r r)e^{i\bar{m}\bar{\theta}}e^{-i \bar{k}_z z}$ in (\ref{TortPressDiff}) yields the dispersion relation
\begin{equation}\label{DispersionOm}
\left(\frac{\bar{\omega}}{c}\right)^2 =\alpha^{-1}_{rr}\bar{k}_r^2 +\alpha^{-1}_{zz} \bar{k}_z^2.
\end{equation}

To proceed further, we must invert the matrix formed by cylindrical components of $\bm{H}$. Fortunately, $\bm{H}$ is block diagonal with respect to the axial ($z$) and transverse ($\{r,\bar{\theta}\}$) coordinates. From the $z$-component (\ref{AxialPr}) of $\nabla p = -\bm{H}\bm{v}$ we immediately have $H^{-1}_{zz}=\left(H_{zz}\right)^{-1}=\left(\frac{\mu}{K_0}-\frac{i\bar{\omega}\rho_0}{\varphi}\right)^{-1}$. The remaining transverse block of $\bm{H}$ will be denoted by $\bm{H}_\perp$, such that
\begin{equation}
\bm{H}_\perp = 
  \begin{pmatrix}
	\frac{\mu}{K_0}-\frac{i\bar{\omega}\rho_0}{\varphi}\left(1+\frac{\Omega^2}{\bar{\omega}^2}\right) & -\frac{2\Omega\rho_0}{\varphi} \\
	\frac{2\Omega\rho_0}{\varphi} & \frac{\mu}{K_0}-\frac{i\bar{\omega}\rho_0}{\varphi}\left(1+\frac{\Omega^2}{\bar{\omega}^2}\right)
	\end{pmatrix}.
\end{equation}
The inverse is given by a similar expression, but with the signs of the off-diagonal elements swapped and an overall factor of $\left(\det{\bm{H}_\perp}\right)^{-1}=\left(H_{rr}^2+H_{r\theta}^2\right)^{-1}$ in front. 

Defining $\bar{\omega}_c=\mu\varphi/\rho_0 K_0$ and $\bar{\omega}_\Omega=\bar{\omega}(1+\Omega^2/\bar{\omega}^2)$, the components of $\bm{\alpha}^{-1}$ relevant to completing specification of the dispersion relation (\ref{DispersionOm}) can be written
\begin{equation}\label{AlphaInvzz}
\alpha_{zz}^{-1}=\frac{\bar{\omega}}{\bar{\omega}+i\bar{\omega}_c}
\end{equation}
and
\begin{equation}\label{AlphaInvrr}
\alpha_{rr}^{-1}=\frac{\bar{\omega}\left(\bar{\omega}_\Omega+i\bar{\omega}_c\right)}{\left(\bar{\omega}_\Omega+i\bar{\omega}_c\right)^2-4\Omega^2}.
\end{equation}

\subsection{Interface Conditions}

Mass conservation at the fluid-disk interface implies continuity of the normal fluid displacement, provided the flow remains laminar. For harmonic acoustic modes at a nonrotating interface, this is equivalent to continuity of normal fluid velocity; allowing the disk to rotate shifts the effective frequency of acoustic modes within the disk, producing a discontinuity in the normal fluid velocity. In \cite{Silke2016}, sound absorption by a rotating porous object is described in the co-rotating frame by an impedance condition $p=-\bar{Z} \, \bm{u}\cdot\bm{n}$, where $\bm{n}$ is the unit surface normal and the co-rotating surface impedance $\bar{Z}$ is a frequency-dependent complex function that encapsulates how the surface interacts with impinging sound waves \cite{Acoustics,Impedance}. For the monochromatic modes under consideration, the co-rotating impedance condition is 
\begin{equation}\label{Bdry}
\left. \frac{\partial_z\Phi}{\Phi}\right|_{z=0}=\frac{-i\rho_0\bar{\omega}}{\bar{Z}} \, ,
\end{equation}
which rests on the assumption that an impedance condition applies in a frame co-rotating with the disk. We now relax this assumption, and by careful analysis of fluid behaviour at the surface of the disk derive an emergent laboratory-frame surface impedance.

The surface of our porous disk is at $z=0$, with the pure-fluid region (hereafter be referred to as ``region I'') occupying the positive $z$ half-space within the tube of radius $R$. To accurately predict how sound scatters from the rotating disk, we must determine how acoustic modes from region I couple at the fluid-disk interface to accessible modes within the porous disk (hereafter referred to as ``region II''). 

The acoustic fluid pressure is continuous across the interface, as expressed by the relation
\begin{equation}\label{Pcont}
p|_{z=0^+}=p|_{z=0^-}\,.
\end{equation}
We denote the co-rotating axial wavenumber by $\bar{k}_z$, the co-rotating radial wavenumber by $\bar{k}_r$, and the co-rotating frequency by $\bar{\omega}$. 

Suppose the disk has thickness $l$, such that both the $e^{-i\bar{k}_z z}$ mode and the $e^{i\bar{k}_z z}$ mode are relevant within the disk. The pressure in region II can then be written as
\begin{equation}
p = J_{\bar{m}}(\bar{k}_r r)e^{i\bar{m}\bar{\theta}}\left(\alpha_+e^{i\bar{k}_z z}+\alpha_-e^{-i\bar{k}_z z}\right)e^{-i\bar{\omega}t}.
\end{equation}
where $\bar{\theta}=\theta-\Omega t$, with $\theta$ being the azimuthal angle defined with respect to the nonrotating (laboratory) coordinate system. In region I, the acoustic fluid velocity field is $\bm{u}=\nabla \Phi$, and the acoustic pressure is $p = i \omega \rho_0 \Phi$, with the potential $\Phi$ given by solutions to the Helmholtz equation of the form
\begin{equation}\label{TotalPotSup}
\Phi=J_m(k_r r)e^{im\theta}\left(\alpha^+ e^{ik_z z}+\alpha^-e^{-ik_z z}\right)e^{-i\omega t}.
\end{equation}

Using a rigid backing for the disk at $z=-l$, which constrains the normal fluid velocity to vanish, one finds
\begin{equation}\label{RigBack}
\alpha_+=\alpha_-e^{2i\bar{k}_z l}.
\end{equation}
Pressure continuity (\ref{Pcont}) then implies $\bar{k}_r=k_r$, $\bar{m}=m$, $\bar{\omega}=\omega-m\Omega$, and
\begin{align}\label{ContPF}
\alpha_-\left(1+e^{2i\bar{k}_z l}\right)=i\omega\rho_0\left(\alpha^++\alpha^-\right).
\end{align} 

In the absence of rotation, continuity of the normal displacement implies continuity of the normal velocity for harmonic modes, since the time dependence appears as $e^{-i\omega t}$. For nonvanishing rotations, acoustic modes that vary as $e^{-i\omega t}$ with respect to the laboratory frame vary as $e^{-i\bar{\omega} t}$ with respect to the co-rotating frame. For these modes, differentiation with respect to time within the disk is thus equivalent to multiplication by $-i\bar{\omega}$, i.e. the relevant frequency relating to time evolution is the \emph{convective} frequency (in this case equal to the co-rotating frequency $\bar{\omega}=\omega-m\Omega$). Consequently, continuity of the normal fluid displacement implies a discontinuity in the normal fluid velocity; for modes that are harmonic with respect to the laboratory frame, the quantity that is continuous across the fluid-disk interface is the fluid velocity divided by the convective frequency. 

Since only a fraction $\varphi$ of the disk volume is occupied by the saturating fluid, one can distinguish between the filtration velocity $\bm{v}$ and the intrinsic fluid velocity $\bm{V}$, obtained by averaging only over pores themselves. The two types of averages are connected by the Dupuit-Forchheimer relationship, $\bm{v}=\varphi \bm{V}$, which expresses quantitatively the difference between averaging the pore-scale fluid velocity over the whole RVE and averaging just over the pores \cite{ConvectionPorousMedia}. Therefore, the correct expression connecting the normal velocities across the interface is
\begin{equation}\label{FreqScaledV}
\frac{1}{-i\omega}\bm{u}\cdot\hat{\bm{z}}|_{z=0^+}=\frac{1}{-i\bar{\omega}}\bm{v}\cdot\hat{\bm{z}}|_{z=0^-}.
\end{equation}

Using (\ref{AxialPr}), the condition (\ref{FreqScaledV}) implies
\begin{equation}\label{ContNormVF}
\frac{-\bar{k}_z\alpha_-\left(1-e^{2i\bar{k}_z l}\right)}{\bar{\omega}\left(\frac{i\bar{\omega}\rho_0}{\varphi}-\frac{\mu}{K_0}\right)}=\frac{k_z}{\omega}\left(\alpha^+-\alpha^-\right).
\end{equation}  
Combining (\ref{ContPF}) and (\ref{ContNormVF}), we find
\begin{equation}
k_z \left(\alpha^+-\alpha^-\right)=-\hat{k}\left(\alpha^++\alpha^-\right),
\end{equation}
with the definition
\begin{equation}
\hat{k}=\frac{i\omega^2 \rho_0 \bar{k}_z K_0\varphi\left(1-e^{2i\bar{k}_zl}\right)}{\bar{\omega}\left(i\bar{\omega}\rho_0 K_0-\mu\varphi\right)\left(1+e^{2i\bar{k}_zl}\right)}=\frac{\omega^2\varphi\bar{k}_z\tanh{\left(-i\bar{k}_zl\right)}}{\bar{\omega}\left(\bar{\omega}+i\bar{\omega}_c\right)}.
\end{equation}
It follows that the reflection amplitude $\alpha^+/\alpha^-$ is given by
\begin{equation}\label{ReflSupF}
\frac{\alpha^+}{\alpha^-}=\frac{k_z-\hat{k}}{k_z+\hat{k}}.
\end{equation}

The surface impedance is defined as $Z=\frac{-p}{\bm{u}\cdot\hat{\bm{n}}}$, with $\hat{\bm{n}}$ being the unit surface normal. In our case $\hat{\bm{n}}=\hat{\bm{z}}$, which leads to the expressions
\begin{equation}\label{ImpedanceSupF}
Z=\frac{\rho_0\omega}{\hat{k}}=\frac{\rho_0\bar{\omega}\left(\bar{\omega}+i\bar{\omega}_c\right)}{\varphi\omega \bar{k}_z\tanh{\left(-i\bar{k}_zl\right)}}.
\end{equation}
The impedance (\ref{ImpedanceSupF}) can also be used to express the reflection amplitude as
\begin{equation}\label{ReflZ}
\frac{\alpha^+}{\alpha^-}=\frac{Z-\frac{\rho_0\omega}{k_z}}{Z+\frac{\rho_0\omega}{k_z}}.
\end{equation}

The appearance of the factor $\tanh{\left(-i\bar{k}_zl\right)}$ in (\ref{ImpedanceSupF}) indicates the existence of points of perfect reflection due to the finite cavity size. These perfect reflection points occur where $\tanh{x}$ either vanishes or has singular points. The former are located at $x=\pi i n$, for $n\in\mathbb{Z}$; at these points $Z$ diverges, but $\alpha^+/\alpha^-=1$. The latter are located at $x=\pi i/2+\pi i n$, for $n\in\mathbb{Z}$; at these points $Z$ vanishes, but $\alpha^+/\alpha^-=-1$.

The radial and axial wavenumbers in region I are given by $k_r$ and $k_z$, and are related to the total wavenumber $k$ by the relation $k^2=k_r^2+k_z^2$. The boundary condition (\ref{BCR}) then implies $J_m'(k_r R)=0$. Denoting the $n^{{th}}$ zero of $J_m'$ by $x_{mn}$, we can equivalently express this condition as
$
k_r R = x_{mn},
$
which makes explicit the constraint imposed on the allowed radial wavenumbers. Therefore, there are two kinds of modes in this setup: propagating modes, where $k_z=\sqrt{k^2-k_r^2}$ is real, and evanescent modes, with purely imaginary $k_z$.

\section{Superradiant amplification}

\subsection{Amplification Conditions}

In our setup the incoming wave travels in the negative $z$ direction, and a reflected wave travels back in the positive $z$ direction. The amplification factor is defined as
$
A_{\omega m}=\frac{|\alpha^+|^2}{|\alpha^-|^2}-1,
$
The amplification factor can be calculated using (\ref{ImpedanceSupF}) and (\ref{ReflZ}). To express $A_{\omega m}$ compactly, we will first rewrite (\ref{ReflZ}) as
\begin{equation}\label{ReflChi}
\frac{\alpha^+}{\alpha^-}=\frac{1-\chi}{1+\chi},
\end{equation} 
with the definition $\chi=\rho_0 \omega/k_z Z$. The quantity $\chi$ is determined by (\ref{ImpedanceSupF}) to be
\begin{equation}
\chi=\frac{\varphi\bar{k}_z \omega^2\tanh{\left(-i\bar{k}_zl\right)}}{k_z\bar{\omega}\left(\bar{\omega}+i\bar{\omega}_c\right)}=\frac{\varphi\bar{k}_z\omega^2\left(\bar{\omega}-i\bar{\omega}_c\right)\tanh{\left(-i\bar{k}_zl\right)}}{k_z\bar{\omega}\left(\bar{\omega}^2+\bar{\omega}_c^2\right)}.
\end{equation} 
The amplification factor can then be written as
\begin{equation}
A_{\omega m}=\frac{-4\text{Re}[\chi]}{|1+\chi|^2}=\frac{-4\text{Re}[\chi^{-1}]}{|\chi\left(1+\chi\right)|^2}.\label{AmpCompact}
\end{equation}
Since the real part of the impedance is positive for absorptive materials, we can deduce from (\ref{AmpCompact}) that positive amplification occurs when the real part of $k_z Z$ is negative. Positive amplification of propagating modes thus occurs when $\text{Re}[Z]<0$. When the incident mode is evanescent, on the other hand, we have $\text{Re}[k_z]=0$ and $\text{Im}[k_z]\equiv\kappa>0$, so superradiance occurs when $\text{Im}[Z]>0$.

Using (\ref{AmpCompact}), the condition for a positive amplification factor can be alternatively expressed as
\begin{equation}\label{GenFinAmp}
Q_l(\bar{\omega})<0,
\end{equation}
with the definition
\begin{align}\label{Q}
&Q_l(\bar{\omega})\equiv\\
&\bar{\omega}\biggl(\bar{\omega}\text{Re}[k_z^*\bar{k}_z\tanh{\left(-i\bar{k}_zl\right)}]+\bar{\omega}_c\text{Im}[k_z^*\bar{k}_z\tanh{\left(-i\bar{k}_zl\right)}]\biggr).\nonumber
\end{align}
The amplification condition (\ref{GenFinAmp}) is valid for either propagating or evanescent incident modes. 

For propagating incident modes ($k_z>0$), the quantity (\ref{Q}) reduces to
\begin{align}\label{PropQ}
&Q_l(\bar{\omega})=\\
&\bar{\omega}k_z\left(\bar{\omega}\text{Re}[\bar{k}_z\tanh{\left(-i\bar{k}_zl\right)}]+\bar{\omega}_c\text{Im}[\bar{k}_z\tanh{\left(-i\bar{k}_zl\right)}]\right).\nonumber
\end{align}
In the evanescent case, for which $k_z=i\kappa$ with $\kappa>0$, (\ref{Q}) becomes
\begin{align}\label{EvFinAmp}
&Q_l(\bar{\omega})=\\
&\bar{\omega}\kappa\left(\bar{\omega}\text{Im}[\bar{k}_z\tanh{\left(-i\bar{k}_zl\right)}]-\bar{\omega}_c\text{Re}[\bar{k}_z\tanh{\left(-i\bar{k}_zl\right)}]\right).\nonumber
\end{align}

Now consider the limit where the disk thickness is large, such that $l\text{Im}[\bar{k}_z]\rightarrow \text{sgn}(\text{Im}[\bar{k}_z])\cdot\infty$. In this case, the ``thick disk'' limit, we have $\tanh{\left(-i\bar{k}_zl\right)}\rightarrow \text{sgn}(\text{Im}[\bar{k}_z])$. It follows that
\begin{equation}
\hat{k}\rightarrow \text{sgn}(\text{Im}[\bar{k}_z])\frac{\varphi \bar{k}_z \omega^2}{\bar{\omega}\left(\bar{\omega}+i\bar{\omega}_c\right)},
\end{equation}
which yields the impedance
\begin{equation}
Z\rightarrow \text{sgn}(\text{Im}[\bar{k}_z])\frac{\rho_0\bar{\omega}\left(\bar{\omega}+i\bar{\omega}_c\right)}{\varphi \bar{k}_z\omega}
\end{equation}
and the amplification parameter
\begin{equation}
\chi\rightarrow \text{sgn}(\text{Im}[\bar{k}_z])\frac{\varphi \bar{k}_z\omega^2 }{k_z\bar{\omega}\left(\bar{\omega}+i\bar{\omega}_c\right)}.
\end{equation}
An approximate superradiance condition for the thick disk limit is therefore given by
\begin{align}\label{ThickSup}
\text{sgn}(\text{Im}[\bar{k}_z])\,\text{Re}\left[\frac{\alpha^{-1}_{zz}\bar{k}_z}{k_z}\right]<0.
\end{align}
For propagating incident modes, (\ref{ThickSup}) implies
\begin{align}\label{ThickSupP}
\text{sgn}(\text{Im}[\bar{k}_z])\,\bar{\omega}\left(\bar{\omega}\text{Re}\left[\bar{k}_z\right]+\bar{\omega}_c\text{Im}\left[\bar{k}_z\right]\right)<0,
\end{align}
but for evanescent incident modes one instead finds 
\begin{align}\label{ThickSupE}
\text{sgn}(\text{Im}[\bar{k}_z])\,\bar{\omega}\left(\bar{\omega}\text{Im}\left[\bar{k}_z\right]-\bar{\omega}_c\text{Re}\left[\bar{k}_z\right]\right)<0.
\end{align}

Shifting attention to the ``thin disk'' limit, in which $l\rightarrow 0$, we have $\tanh{\left(-i\bar{k}_zl\right)}\sim -i\bar{k}_zl+\mathcal{O}(l^3)$. This implies
\begin{equation}
\hat{k}\sim -\frac{i\varphi \omega^2 l\bar{k}_z^2}{\bar{\omega}\left(\bar{\omega}+i\bar{\omega}_c\right)}+\mathcal{O}(l^3).
\end{equation}  
One can then observe that the impedance $Z=\rho_0\omega/\hat{k}$ diverges as $l^{-1}$ when $l\rightarrow 0$, while the amplification parameter $\chi$ behaves as 
\begin{equation}
\chi\sim -\frac{i\varphi l\bar{k}_z^2\omega^2}{k_z\bar{\omega}\left(\bar{\omega}+i\bar{\omega}_c\right)}+\mathcal{O}(l^3).\label{ThinAmp}
\end{equation}
Superradiance requires $\text{Re}[\chi]<0$; using (\ref{ThinAmp}) and the relation $\alpha_{zz}^{-1}=\bar{\omega}/(\bar{\omega}+i\bar{\omega}_c)$, we obtain
\begin{equation}
\text{Re}\left[\chi\right]\sim -\frac{\varphi l \omega^2}{\bar{\omega}^2}\text{Re}\left[\frac{i\alpha_{zz}^{-1}\bar{k}_z^2}{k_z}\right]+\mathcal{O}(l^3).
\end{equation}
Thus, an approximate superradiance condition for the thin disk limit is given by
\begin{align}\label{ThinSup}
\text{Re}\left[\frac{i\alpha^{-1}_{zz}\bar{k}_z^2}{k_z}\right]>0\hspace{5pt}\Leftrightarrow\hspace{5pt} \text{Im}\left[\frac{\alpha^{-1}_{zz}\bar{k}_z^2}{k_z}\right]<0.
\end{align}
For propagating incident modes, (\ref{ThinSup}) implies
\begin{align}\label{ThinSupP}
\bar{\omega}\left(\bar{\omega}_c\text{Re}\left[\bar{k}_z^2\right]-\bar{\omega}\text{Im}\left[\bar{k}_z^2\right]\right)>0,
\end{align}
while for evanescent incident modes one finds
\begin{align}\label{ThinSupE}
\bar{\omega}\left(\bar{\omega}\text{Re}\left[\bar{k}_z^2\right]+\bar{\omega}_c\text{Im}\left[\bar{k}_z^2\right]\right)>0.
\end{align}

\subsection{Superradiance}

We now show that propagating modes can superradiate, using the general amplification condition (\ref{GenFinAmp}). To do so, it suffices to show that $Q_l$ (defined by (\ref{Q})) changes sign as $\bar{\omega}$ crosses the origin. The function $\tanh(z)$ is analytic and odd; provided $|z|<\pi/2$, we have the following convergent series representation
\begin{equation}\label{TanhSer}
\tanh(z)=\sum_{n=1}^{\infty}\frac{2^{2n}\left(2^{2n}-1\right)B_{2n}z^{2n-1}}{\left(2n\right)!},
\end{equation}
with $B_{2n}$ denoting the Bernoulli numbers. It follows from (\ref{DispersionOm}) that $\bar{k}_z^2$ vanishes at $\bar{\omega}=0$; for sufficiently small $\bar{\omega}$, the representation (\ref{TanhSer}) then implies that $\bar{k}_z\tanh{\left(-i\bar{k}_zl\right)}$ has an expansion involving only even powers of $\bar{k}_z$. From (\ref{DispersionOm}), we can also deduce that when $\bar{\omega}\rightarrow-\bar{\omega}$ (with $\Omega$ held constant), we have $\bar{k}_z^2\rightarrow\left(\bar{k}_z^2\right)^*$. Therefore, we also have
\begin{equation}
\bar{k}_z\tanh{\left(-i\bar{k}_zl\right)}\rightarrow -\left[\bar{k}_z\tanh{\left(-i\bar{k}_zl\right)}\right]^*,
\end{equation}
and consequently
\begin{equation}
Q_l\rightarrow -Q_l,
\end{equation}
which completes our demonstration that superradiance can occur for propagating incident modes.

The same reasoning used in the propagating case leads to the conclusion that as $\bar{\omega}$ crosses the origin, the expression (\ref{EvFinAmp}) for $Q_l$ does not change sign. However, this does not mean evanescent incident modes cannot superradiate, for the following reason: the transformation $\bar{\omega}\rightarrow-\bar{\omega}$ considered above treated $\bar{\omega}$ and $\Omega$ as independent variables, such that $\Omega$ was held constant. Alternatively, the transformation $\bar{\omega}\rightarrow-\bar{\omega}$ can be made with $\omega$ and $m$ held constant (take $m=1$, for simplicity). The induced $\bar{k}_z^2$ transformation then differs from the previous $\bar{k}_z^2\rightarrow\left(\bar{k}_z^2\right)^*$, due to the explicit $\Omega$ dependence of $\alpha_{rr}^{-1}$ in the dispersion relation (\ref{DispersionOm}). 

Specifically, since for $m=1$ we have $\bar{\omega}=\omega-\Omega$, the alternate transformation $\bar{\omega}\rightarrow-\bar{\omega}$ with $\omega$ held constant implies $\Omega=\omega-\bar{\omega}\rightarrow\omega+\bar{\omega}=\Omega+2\bar{\omega}$. Using the notation $\tilde{\omega}\equiv\bar{\omega}+i\bar{\omega}_c$ and $\tilde{\omega}_+\equiv\bar{\omega}_\Omega+i\bar{\omega}_c=\tilde{\omega}+\Omega^2/\bar{\omega}$, we have $\tilde{\omega}\rightarrow-\tilde{\omega}^*$, $\tilde{\omega}_+\rightarrow-\tilde{\omega}_+^*-4\omega$, and
\begin{align}
\alpha_{rr}^{-1}=\frac{\bar{\omega}\tilde{\omega}_+}{\tilde{\omega}_+^2-4\Omega^2}\rightarrow&\frac{\bar{\omega}\left(\tilde{\omega}_+^*+4\omega\right)}{\left(\tilde{\omega}_+^*+4\omega\right)^2-4\Omega^2-16\bar{\omega}\omega}\\
&\neq \left(\alpha_{rr}^{-1}\right)^*=\frac{\bar{\omega}\tilde{\omega}_+^*}{\left(\tilde{\omega}_+^*\right)^2-4\Omega^2}.
\end{align}
In these expressions, $\Omega$ is used to denote $\omega-\bar{\omega}$, since $\omega$ is taken to be a fixed quantity.

To prove that evanescent modes can indeed superradiate, we restrict our attention to the thick disk limit ($\l\rightarrow\infty$), defined above. We first re-express the thick disk amplification condition (\ref{ThickSupE}) as
\begin{equation}\label{ThickSupE1}
Q_\infty(\bar{\omega})<0,
\end{equation}
for 
\begin{align}\label{ThickSupE2}
Q_\infty(\bar{\omega})\equiv\bar{\omega}\left(\bar{\omega}\,\left|\,\text{Im}\left[\bar{k}_z\right]\right|-\text{sgn}(\text{Im}[\bar{k}_z^2])\bar{\omega}_c\,\left|\,\text{Re}\left[\bar{k}_z\right]\right|\right),
\end{align}
From the form of (\ref{ThickSupE2}), it is clear that the evanescent mode amplification condition (\ref{ThickSupE1}) is independent of the choice of branch cut used to define the wavevector $\bar{k}_z$. The quantities $\left|\,\text{Re}\left[\bar{k}_z\right]\right|$ and $\left|\,\text{Im}\left[\bar{k}_z\right]\right|$ appearing in (\ref{ThickSupE2}) can be written explicitly as
\begin{align}\label{Rekzbar}
\left|\,\text{Re}\left[\bar{k}_z\right]\right|=\sqrt{\frac{|\bar{k}_z^2|+\text{Re}[\bar{k}_z^2]}{2}}
\end{align}
and
\begin{align}\label{Imkzbar}
\left|\,\text{Im}\left[\bar{k}_z\right]\right|=\sqrt{\frac{|\bar{k}_z^2|-\text{Re}[\bar{k}_z^2]}{2}},
\end{align}
where $\bar{k}_z^2$ is obtained from the dispersion relation (\ref{DispersionOm}). The inverse tortuosity components $\alpha_{zz}^{-1}$ and $\alpha_{rr}^{-1}$ are specified by (\ref{AlphaInvzz}) and (\ref{AlphaInvrr}), respectively.

Using the limiting behaviour of the dispersion relation (\ref{DispersionOm}),  we can now show that it must be possible to satisfy (\ref{ThickSupE1}). Recall that $\bar{k}_z^2=0$ when $\bar{\omega}=0$, which implies that $Q_\infty(0)=0$. One can then deduce from
\begin{equation}
\left.\left(\frac{\partial \bar{k}_z^2}{\partial \bar{\omega}}\right)\right|_{\bar{\omega}=0}=\frac{i \bar{\omega}_c}{\omega^2}\left[\left( \frac{\omega}{c}\right)^2-k_r^2\right]
\end{equation}
and 
\begin{equation}
\left.\left(\frac{\partial^2 \bar{k}_z^2}{\partial \bar{\omega}^2}\right)\right|_{\bar{\omega}=0}=\frac{2}{\omega^2}\left[\left(\frac{\omega}{c}\right)^2-k_r^2\left(1+\frac{\bar{\omega}_c^2}{\omega^2}\right)\right]-\frac{4i\bar{\omega}_ck_r^2}{\omega^3}
\end{equation}
that for sufficiently small $\bar{\omega}>0$, we must have $\text{Re}[\bar{k}_z^2]<0$ and $\text{Im}[\bar{k}_z^2]<0$. Setting $\text{sgn}(\text{Im}[\bar{k}_z^2])=-1$ in (\ref{ThickSupE2}) then yields $Q_\infty>0$, indicating no amplification (as expected).

For sufficiently small $\bar{\omega}<0$, it can be similarly shown that $\text{Re}[\bar{k}_z^2]<0$ and $\text{Im}[\bar{k}_z^2]>0$. We then have $\text{sgn}(\text{Im}[\bar{k}_z^2])=+1$, but since $\bar{\omega}$ has changed sign, we obtain from (\ref{ThickSupE2}) the unexpected result $Q_\infty>0$, again indicating no amplification. Actually, we can see directly from (\ref{ThickSupE1}) that for any $\bar{\omega}<0$, $Q_\infty$ cannot be negative unless $\text{Im}[\bar{k}_z^2]$ is negative. At the very least, then, there must exist an interval of $\bar{\omega}<0$ (with $\bar{\omega}=0$ as a limit point) for which thick disks cannot superradiate evanescent modes, despite satisfying the usual superradiance condition $\bar{\omega}<0$.

Fortunately, things change as $\bar{\omega}$ becomes more negative. Assuming that $\bar{\omega}_c\neq 0$, if we take $\bar{\omega}\rightarrow-\infty$ while keeping $\omega$ fixed, the dispersion relation (\ref{DispersionOm}) implies
\begin{align}
\text{Re}\left[ \bar{k}_z^2\right]\sim\left(\frac{\bar{\omega}}{c}\right)^2\rightarrow\infty, \\
\text{Im}\left[ \bar{k}_z^2\right]\sim\bar{\omega}\left(\frac{\bar{\omega}_c}{c^2}+\frac{k_r^2}{2\bar{\omega}_c}\right)\rightarrow-\infty.
\end{align}
Both $\text{Re}\left[ \bar{k}_z^2\right]$ and $\text{Im}\left[ \bar{k}_z^2\right]$ must therefore switch signs at some point. More precisely, there must exist a pair of distinct negative $\bar{\omega}$ values, denoted by $\bar{\omega}_{R}$ and $\bar{\omega}_{I}$, such that $\text{Re}\left[ \bar{k}_z^2\right](\bar{\omega}_{R})=0$, $\text{Im}\left[ \bar{k}_z^2\right](\bar{\omega}_{I})=0$, $\text{Re}\left[ \bar{k}_z^2\right](\bar{\omega}<\bar{\omega}_{R})>0$, and $\text{Im}\left[ \bar{k}_z^2\right](\bar{\omega}<\bar{\omega}_{I})<0$.

Now take $\bar{\omega}<\text{min}\{\bar{\omega}_{R},\bar{\omega}_{I}\}\equiv \bar{\omega}_*$. In this case we have $\text{Re}\left[ \bar{k}_z^2\right]>0$ and $\text{Im}\left[ \bar{k}_z^2\right]<0$. Using expressions (\ref{Rekzbar}) and (\ref{Imkzbar}), one readily obtains 
$\left|\,\text{Re}\left[\bar{k}_z\right]\right|>\left|\,\text{Im}\left[\bar{k}_z\right]\right|$, and it follows that 
\begin{align}
Q_\infty=&\,\left|\bar{\omega}^2\text{Im}\left[\bar{k}_z\right]\right|-\bar{\omega}_c\left|\,\bar{\omega}\text{Re}\left[\bar{k}_z\right]\right|\nonumber\\
<&\,\left|\bar{\omega}\right|\left(|\bar{\omega}|-\bar{\omega}_c\right).\label{EvSuper}
\end{align}
Thus, one can obtain the desired result $Q_\infty<0$ (i.e. amplification), provided one can show that $\bar{\omega}_c>|\bar{\omega}_*|$. Accordingly, we now investigate how various frequency scales affect the zeros of $\text{Re}\left[\bar{k}_z^2\right]$ and $\text{Im}\left[\bar{k}_z^2\right]$.

In what follows, it will be useful to keep in mind the nonrotating limit ($\Omega=0$); in this case $\bar{\omega}=\omega$, and the dispersion relation (\ref{DispersionOm}) reduces to
\begin{equation}
\bar{k}_z^2=\left(\frac{\omega}{c}\right)^2-k_r^2+\frac{i\bar{\omega}_c\omega}{c^2}.
\end{equation}
Hereafter, we take the rotation rate $\Omega$ to be positive, which implies that $\bar{\omega}<\omega$. We also find a bound on $\bar{\omega}$ that arises if we wish to keep the tangential velocity at the outer edge of the disk lower than the speed of sound: $\omega-\bar{\omega}=\Omega<c/R$.

Another notable reference is $\bar{\omega}=-\omega$, which corresponds to
\begin{equation}\label{negomeg}
\bar{k}_z^2=\left[\left(\frac{\omega}{c}\right)^2-k_r^2 +\frac{28 k_r^2 \omega^2}{49\omega^2+\bar{\omega}_c^2}\right]-\frac{i\bar{\omega}_c\omega}{c^2}\left[1+\frac{4 c^2 k_r^2}{49\omega^2+\bar{\omega}_c^2}\right].
\end{equation}
The significance of (\ref{negomeg}) is clear, since $\text{Im}\left[\bar{k}_z^2\right]$ is manifestly negative. The sign of $\text{Re}\left[\bar{k}_z^2\right]$, on the other hand, depends on the relative sizes of $\omega$, $\bar{\omega}_c$, and $c\cdot k_r$. The incident mode is evanescent, so we can write $\omega=\epsilon_k c\cdot k_r$, for some $\epsilon_k \in (0,1)$. To obtain $Q_\infty<0$, according to (\ref{EvSuper}), we must have $|\bar{\omega}|=\epsilon_c \bar{\omega}_c$, for some $\epsilon_c \in (0,1)$. Direct computation using (\ref{negomeg}) then leads to the conclusion that by choosing $\epsilon_k$ close to $1$, $\text{Re}\left[\bar{k}_z^2\right]>0$ for a wide range of $\epsilon_c \in (0,1)$; in this range, the system exhibits superradiance.

The preceding considerations imply that there must exist a point $\bar{\omega}_*<0$ for which $Q_\infty$ changes sign; it is this point that signals the transition to a superradiant regime for evanescent incident modes, in the thick disk limit. No amplification occurs in the interval $\bar{\omega}_*<\bar{\omega}<0$ - we can therefore interpret this as a band gap. 

\begin{figure}[t!]
\includegraphics[width=0.45\textwidth]{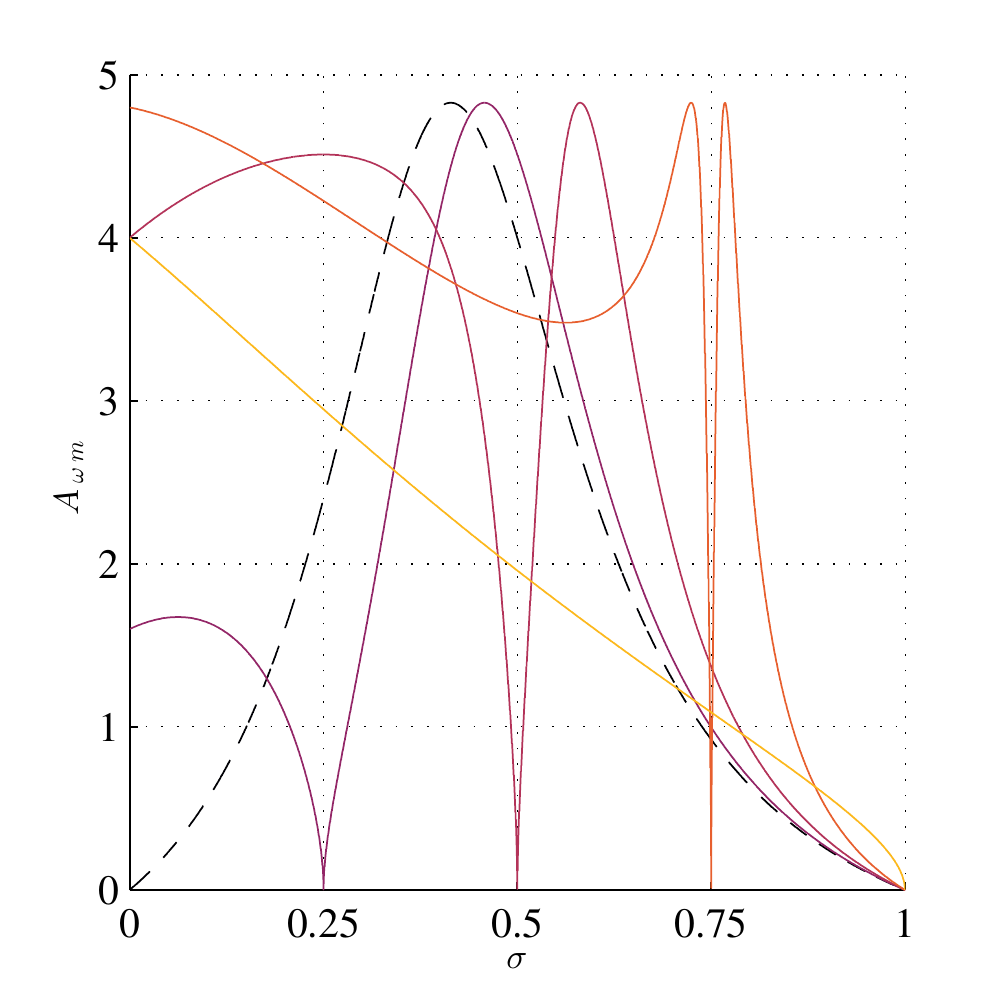} %
\caption{Plot of the amplification factor for acoustic vortex beam superradiance with different choices of $\sigma_\star$ in the interval $[0,1]$, assuming $\hat{Z}=1-i$. The cusp located at $\sigma=\sigma_\star$ indicates where in parameter space the initial wave switches from being evanescent ($\sigma<\sigma_\star$) to propagating ($\sigma>\sigma_\star$).}
	\label{fig:Amp1}
\end{figure}

\section{Experimental Implementation}
We will now discuss a specific experimental proposal for observing superradiance with OAM beams focussing on its feasibility using current technology. The geometrical arrangement of the experimental components is apparent from the schematics in Figure~\ref{fig:Schem}: OAM beam source, an effectively lossless cylindrical enclosure of given inner radius $R$ and length $L$, and a sufficiently rapidly rotating disk composed of sound absorbing material.

For clarity of presentation, we restrict our attention to the approximate impedance condition (\ref{Bdry}); deviations occurring due to the more accurate impedance (\ref{ImpedanceSupF}) are of interest in their own right, and can be explored further upon implementation. The amplification factor can then be written as
\begin{equation}\label{Amp1}
A_{\omega m} = \frac{-4 \rho_0 \bar{\omega}\text{Re}[k_z \bar{Z}]}{|\rho_0 \bar{\omega}+k_z \bar{Z}|^2}  
=\frac{-4 \text{Re}[\bar{\chi}]}{|1+\bar{\chi}|^2},
\end{equation}
with $\bar{\chi}=\rho_0 \bar{\omega}/k_z \bar{Z}$. To plot amplification spectra arising within our setup, we define the following dimensionless quantities: $\hat{Z}=\bar{Z}/\rho_0 c$, $\sigma=\omega/m\Omega$, and 
$\sigma_\star=cx_{mn}/m\Omega R$. It follows that
\begin{equation}
\bar{\chi} = \frac{\left(\sigma-1\right)}{\hat{Z}\sqrt{\sigma^2-\sigma_\star^2}}.
\end{equation}
For co-rotating ($m>0$) modes, the parameter $\sigma_\star$ is positive, and provides a measure of the sound speed relative to the tangential velocity at the outer edge of the disk. The value $\sigma_\star=0$ represents the limit of large disk radius, whereby the axial wavevector $k_z$ is much greater than the radial wavevector $k_r$. The dimensionless impedance of ``typical" porous/fibrous materials is often taken to be $\hat{Z}=1-i$ \cite{Silke2016}. As can be deduced from the complex structure of the full impedance (\ref{ImpedanceSupF}), the accuracy of this estimate depends on the frequency of the acoustic mode (as well as other system parameters).

Figure \ref{fig:Amp1} shows the amplification varying with $\sigma$ for the typical impedance for various values of the dimensionless parameter $\sigma_\star$. When $0\leq\sigma_\star\leq 1$, a cusp forms along the horizontal axis at $\sigma=\sigma_\star$, in which case the axial wavevector vanishes ($k_z=0$) and all of the mode energy comes from angular momentum ($k=k_r$). The value $\sigma_\star$ separates the evanescent ($\sigma<\sigma_\star$) and propagating ($\sigma>\sigma_\star$) regions of the parameter space.
One can also observe from Figure \ref{fig:Amp1} that maxima of the amplification factor have the same height for $0\leq\sigma_\star\leq 1$. The maximal amplification for this entire parameter range can then be determined by considering the $\sigma_\star=0$ case (the $R\rightarrow\infty$ limit), from which one easily obtains a maximal amplification of $A_{\omega m}=\sqrt{2}(2+\sqrt{2})$. Upon comparison with the original superradiance setup \cite{Silke2016}, we find that the new scattering configuration produces a nontrivial increase in the amount of superradiant amplification over much of the parameter space. An exception to this increase occurs for supersonic rotations, but only when the incident waves have very finely-tuned frequencies. For positive values of $\sigma_\star$ that are sufficiently less than unity, we also find a much broader amplification maximum than in the original approach.

Depending on the context, these types of waves are often referred to as either vortex beams, or in the specific case of acoustics, as ``sonic screwdrivers'', due to the torque they can exert on sound absorbing objects \cite{SonicScrew}. OAM beam generation can be accomplished in two different ways. A \textit{passive} technique, using either a spiral grating \cite{BesselBeam2015,SpiralCoil}, acoustic resonance \cite{AcousticResonance}, or a metamaterial \cite{Metasurf}. The passive techniques are relatively easy to construct and require less computational power, but this often comes at the cost of tunability, which would restrict exploration of the parameter space. This in contrast to the \textit{active} technique, using a phase-controlled array of piezoelectric transducers \cite{SonicScrew,TransducArray,CharacterizationPhasedArray}. If one seeks to test the accuracy of the impedance condition for a wide range of beam frequencies and rotation rates, a transducer array is the ideal method of beam generation.

The cylindrical tubing serves to isolate the OAM modes from environmental noise sources, as well as providing infrastructure to monitor the acoustic pressure field.  
The measurement of the amplification factor can be achieved by attaching an array of microphones reaching the inner surface of the tube, which allows the pressure field to be reconstructed. The amplification factor can then be deduced by fitting the acoustic velocity potential $\Phi$ obtained from the acoustic pressure measurements via $\Phi=p/i\omega \rho_0$ with the predicted form (\ref{TotalPotSup}). 
The phase speed $c$ of acoustic waves in air is $343$ m/s, and hence a characteristic length scale for both propagating and evanescent modes is given by $c/f$, where $f$ is its frequency. Therefore, for our system to contain a manageable number of characteristics lengths, it is desirable to work in the kHz frequency range, which corresponds to a length scale of $\lesssim 1$m.
The radial confinement of the tube is limited by the fact that one end of the tube is closed by a rapidly rotating disk of the same radius. 

\begin{figure}
  \includegraphics[width=0.45\textwidth]{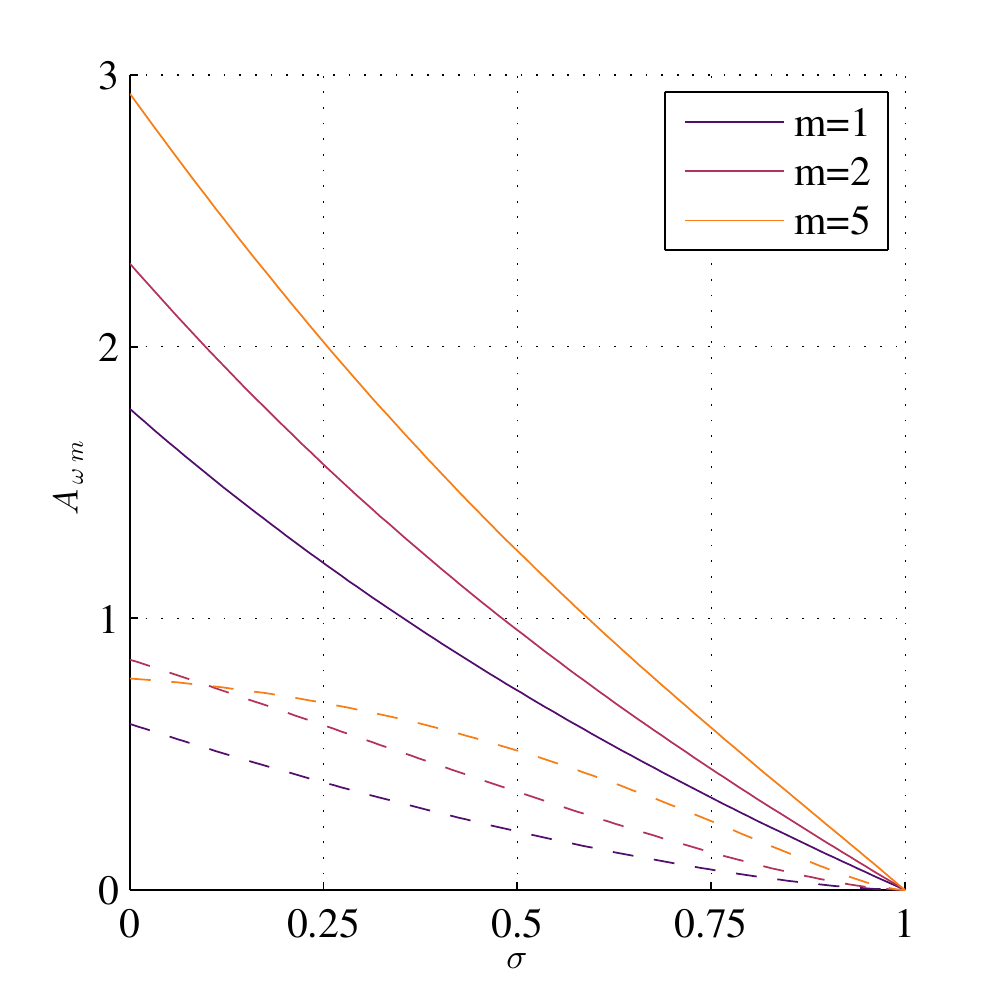}
  \caption{Plot of the amplification factor for evanescent modes, including frequency variation in the surface impedance. Parameters: $\sigma_0 = 40 \text{g}/\text{s}\text{cm}^3$, $R=8\text{cm}$, $\rho = 0.00129 \text{g}/\text{cm}^3$, $c=34300\text{cm}/\text{s}$, $\Omega/2\pi = 675 \text{Hz}$, $n=1$, and $m\in\{1,2,5\}$. The solid lines use the dimensionless impedance $Z=1-i$. The dashed lines use empirical impedance functions \cite{Impedance}, and are valid for $\sigma<0.41$ ($m=1$), $\sigma<0.70$ ($m=2$), and $\sigma<0.88$ ($m=5$).}
	\label{fig:Amp2}
\end{figure}

Beam frequencies any higher than the $\text{kHz}$ range are undesirable, since the superradiance condition implies that we must rotate our porous/fibrous disk at approximately the same frequency, and higher rotation rates are difficult to stabilize. Due to drag, rapid rotations can also create centrifugal pumping, but this effect can be minimized by using a frame with cylindrical pores aligned with the rotation axis of the disk.
Working with rotation rates near the $\text{kHz}$ range requires an extremely strong material, since at the outer disk edge the centripetal acceleration ($a_c=R\Omega^2$) will be approaching $10^5 g$ (with $g=9.8\text{m}\hspace{2pt}\text{s}^{-2}$ being the acceleration due to gravity). Typical porous/fibrous materials cannot sustain the associated tensile loads, but a foam made from the less familiar silicon carbide appears to be able to accommodate our demands \cite{Foams,MatAcoustHandbook,PorousHandbook}. If $\bar{a}_{c}$ is the maximum  centripetal acceleration a specific disk can support, then for subsonic disk motion we must keep the rotation rate $\Omega$ below $\bar{a}_{c}/c$. 

Though superradiance research has historically involved only propagating modes, we consider it preferable experimentally to work with the evanescent modes, for the following reason. Using the relation $k=\omega/c$ and the radial constraint $k_r R = x_{mn}$ we can write the axial wavevector as $k_z=(1/c)\sqrt{\omega^2-(c x_{mn}/R)^2}$, which is real whenever $\omega>c x_{mn}/R$. In the superradiant regime $\omega-m\Omega<0$, we then find that $R\Omega>c(x_{mn}/m)$ for propagating modes. The zeros of $J_m'$ obey $x_{mn}/m>1$, so we conclude that the tangential velocity $R\Omega$ at the outer edge of the disk must exceed the speed of sound $c$. Thus, part of the disk must move supersonically in order to superradiate propagating modes. The shock waves produced by supersonic disk motion make the background flow deviate from its initially irrotational state. Angular momentum then convects away from the disk surface, spinning up the fluid and eventually preventing the occurrence of rotational superradiance altogether. Instabilities that can occur in the original Zel'Dovich configuration are by comparison less problematic, as they do not hinder the measurement of rotational superradiance~\cite{Silke2016}. It may be possible to reduce the generation of shock waves by placing a membrane above the disk surface that transmits acoustic modes but prevents passage of fluid flow; however, by working solely with evanescent modes these complications can be avoided.

Measuring superradiance in the evanescent case requires a different approach than in the propagating case: if $k_z=i\kappa$,  then the squared-norm of (\ref{TotalPotSup}) is proportional to $\cosh{2\kappa (z-z_0)}+\cos{2\alpha}$, with $\alpha^+/\alpha^-=e^{2\kappa z_0 + 2i\alpha}$ ($\alpha$ is a phase-shift, and $z_0$ determines the amplification). By relating the acoustic potential $\Phi$ to the acoustic pressure $p$, one can locate the minimum $z_0$ in the pressure field within the tube to obtain the amplification. 

In practice, the impedance of a sound-absorbing material varies with the frequency of the incident wave. To test whether such frequency dependence affects the preceding results, we consider the empirical impedance function for the porous/fibrous absorbent materials studied in \cite{Impedance}, which has the complex power-law form
$
\hat{Z}\approx 1+9.08\left(f/\sigma_0\right)^{-0.75}  - 11.9\, i\left(f/\sigma_0\right)^{-0.73} \, .
$
This empirical impedance is a function of the frequency $f=\omega/2\pi$ scaled by the flow resistance $\sigma_0$, and is valid for $f$ in the range $250-4000\hspace{3pt}\text{Hz}$, for materials with flow resistances of $2-80\hspace{3pt}\text{g}\hspace{2pt} \text{s}^{-1} \text{cm}^{-3}$ (i.e. $2000-80,000\text{Pa}\hspace{2pt}\text{s}\hspace{2pt}\text{m}^{-2}$) \cite{Impedance}. We plot the resulting superradiant spectra for evanescent modes in Figure \ref{fig:Amp2}. While superradiance is slightly suppressed compared to the idealized case $\hat{Z}=1-i$, the amplification shown in Figure \ref{fig:Amp2} is still larger than the amplification obtained for subsonic rotations in the original configuration \cite{Silke2016}.\\

\section{Conclusion}
We have seen in the example of acousto-mechanics that our new direction for rotational superradiance leads to a simplified theoretical description. The associated effective dimensional reduction of the scattering arrangement also lowers the complexity of the experimental setup. We also derived a more accurate expression for the surface impedance of the fluid-disk interface at rapid relative rotation. 

The results presented above indicate that our proposal offers a promising new way to explore superradiance experimentally. By following our suggested alignment of the wave propagation direction and disk rotation axis the amplification process is more efficient, leading to higher amplification compared to the standard approach. Guided by the theoretical analysis detailed above, our proposal also offers a way to explore uncharted regimes in the acoustics of rapidly-rotating porous media, which would further our understanding of the role angular momenta play in wave-structure interactions. Based on the similarities between acousto-mechanics and optomechanics \cite{AcousticTorque}, experimental progress coupling rotating mirrors to laser modes with very high OAM \cite{Allen97,EffHighTopoCharge,SmallBeamHighTopo,Anton10000} could allow our new direction to be implemented electromagnetically. Recent optomechanics success measuring rovibrational entanglement \cite{RovibeEntangle2008,RotoMirrorEntangle2008} even hints at the possibility of one day observing quantum features of rotational superradiance in a laboratory setting.

\section{Acknowledgements}

The authors thank Jorma Louko, Zack Fifer, August Geelmuyden, and Sir Roger Penrose, for stimulating discussions. The research of WGU is supported by NSERC (Natural Science and Engineering Research Council) of Canada, and also by CIfAR. SW acknowledges financial support provided under the Paper Enhancement Grant at the University of Nottingham, the Royal Society University Research Fellow (UF120112), the Nottingham Advanced Research Fellow (A2RHS2), the Royal Society Project (RG130377) grants, the Royal Society Enhancement Grant (RGF/EA/180286) and the EPSRC Project Grant (EP/P00637X/1). SW acknowledges partial support from STFC consolidated grant No. ST/P000703/. The fellowship held by CG while this research was conducted was funded by NSERC through WGU, with partial support from SW.

\end{document}